\documentclass[aps,prl,twocolumn,showpacs,preprintnumbers,superscriptaddress,floatfix,amsmath,amssymb,amsfonts]{revtex4-1}
\usepackage{graphicx}
\usepackage{dcolumn}
\usepackage[bookmarks=true,colorlinks,linkcolor=blue,urlcolor=blue,citecolor=blue]{hyperref}
\usepackage{color}
\usepackage{subfigure}

\usepackage{epsfig}
\usepackage{amsmath, amssymb}
\usepackage{verbatim}
\usepackage{bm}
\usepackage{mathrsfs}

\makeatletter
\@tfor\next:=abcdefghijklmnopqrstuvwxyzABCDEFGHIJKLMNOPQRSTUVWXYZ\do{%
  \def\command@factory#1{%
    \expandafter\def\csname bf#1\endcsname{\mathbf{#1}}
  }
 \expandafter\command@factory\next
}
\@tfor\next:=abcdefghijklmnopqrstuvwxyzABCDEFGHIJKLMNOPQRSTUVWXYZ\do{%
  \def\command@factory#1{%
    \expandafter\def\csname cal#1\endcsname{\mathcal{#1}}
  }
 \expandafter\command@factory\next
}
\makeatother

\newcommand{\ket}[1]{\left|#1\right>}
\newcommand{\bra}[1]{\left<#1\right|}
\newcommand{\braket}[1]{\left<#1\right>}
\newcommand{\para}[1]{\left(#1\right)}
\newcommand{\abs}[1]{\left|#1\right|}

\newcommand{\COMMENT}[1]{}

\begin{document}

\title{$Z_3$ generalization of the Kitaev's spin-1/2 model}
\author{Abolhassan Vaezi}
\affiliation{Department of Physics, Cornell University, Ithaca NY 14850}

\begin{abstract}
We generalize the Kitaev's spin-1/2 model on the honeycomb by introducing a two-dimensional $Z_3$ clock model on the triangular lattice with three body interaction. We discuss various properties of this model and show that the low energy theory of the $Z_3$ generalized Kitaev model (GKM) is described by a single $Z_3$ parafermion per lattice site coupled to a $Z_3$ gauge field. We also introduce a slave-fermion approach for this GKM, treat the resulting fermionic Hamiltonian at the mean-field level, solve the mean field parameters self-consistently, and obtain the low energy effective Chern-Simons (CS) gauge theory. The resulting CS gauge theory is identical to that of a $(221)$ fractional quantum Hall state. We then go beyond the mean-field approximation and demonstrate that fluctuations generate a uniform interlayer pairing for the dual $(221)$ bilayer state. We argue that this perturbed system can undergo a phase transition to the Fibonacci phase by tuning the interlayer pairing strength. 
\end{abstract}
\maketitle


\section{Introduction} 

Kitaev's honeycomb model \cite{kitaev2006,nussinov2013,Knolle2014} is one of the few examples of the exactly solvable models in theoretical condensed matter physics. This model exhibits a stable $Z_2$ spin liquid phase with non-Abelian excitation in its B-phase. Kitaev showed that after perturbing his model in the B-phase with time reversal breaking perturbations, $Z_2$ vortices will bind single Majorana zero modes. Kitaev also established a mapping between his model and a $p_x+ip_y$ superconductor of spinless neutral fermions coupled to a $Z_2$ gauge field. In this duality transformation, the Majorna zero modes are bound to the vortices of the dual $p_x+ip_y$ superconductor in its weak pairing phase\cite{read2000}. After recent interests in finding fractional topological superconductors\cite{vaezi2013,vaezi2013b,mong2013,vaezi2013c} and related systems \cite{vaezi2014,barkeshli2012a,clarke2013,lindner2012,cheng2012,barkeshli2013genon} with different types of non-Abelian excitations, a natural question that arises is whether we can generalize the Kitaev's model such that (1) its low energy is described by a stable spin liquid coupled to a discrete gauge symmetry and (2) is dual to a fractional topological superconductor with non-Abelian anyons capable of making universal quantum computation through braiding operations. In this paper we give an affirmative answer to these questions and introduce a $Z_3$ generalization of the Kitaev model with a stable $Z_3$ fractionalized spin liquid ground-state. More importantly we will argue that for a wide range of parameters this spin liquid phase belongs to the Fibonacci phase \cite{mong2013,vaezi2013b,nayak2008,trebst2009,vaezi2014}. 

Recently, Barkeshli et al. ~\cite{BarkeshliGKM} have introduced the most direct generalization of the Kitaev's model by replacing spin 1/2 operators with $Z_n$ clock operators \cite{fradkin1980,fendley2012}. The resulting Hamitlonian has many interesting properties similar to the Kitaev's original model. Here we introduce a different generalization of the Kitaev's model that is more tractable than that of Ref.~\cite{BarkeshliGKM} and from which we gain a fair understanding of the two dimensional (2D) parafermion systems as well. We show that the low energy theory of this model is identical to that of a (221) bilayer quantum Hall state with interlayer pairing added to it which is believed to undergo a phase transition to the Fibonacci phase \cite{mong2013,vaezi2013b,nayak2008,trebst2009,vaezi2014}. We finally present another related model with similar properties.
\begin{figure}
\centering
{\includegraphics[width=0.8\columnwidth]{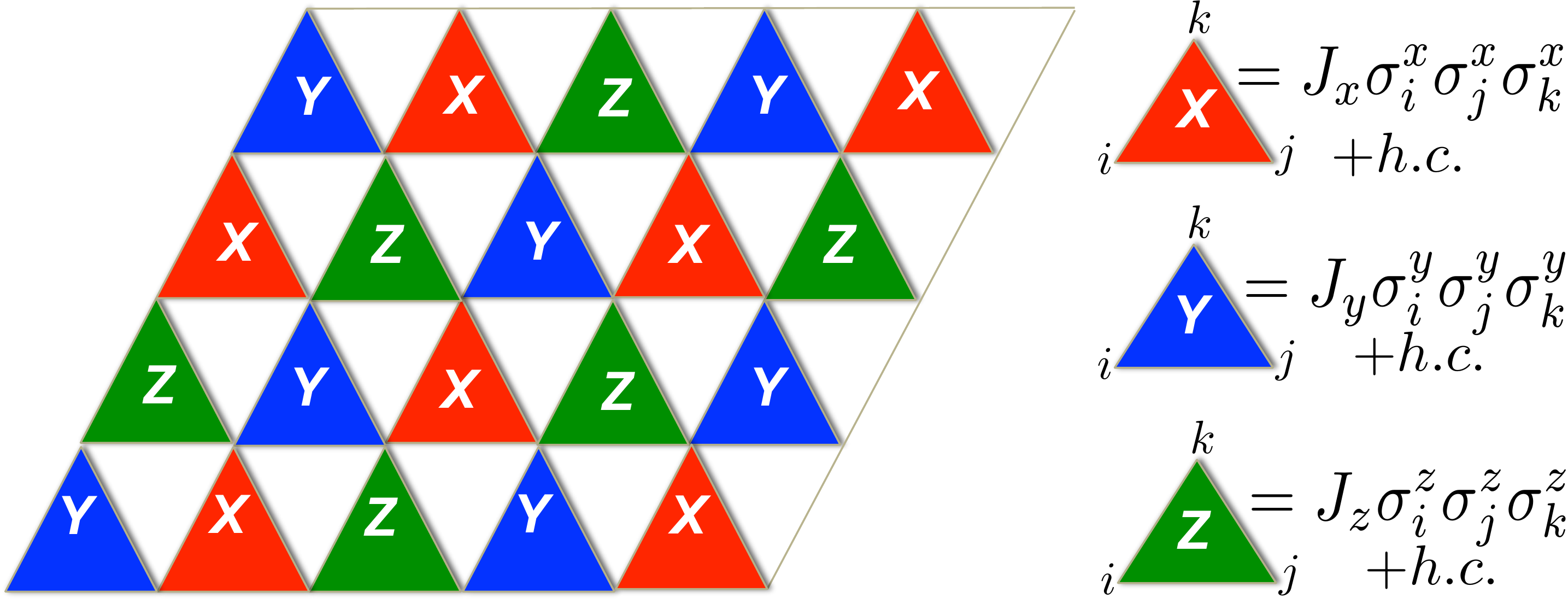}}
\caption{The $Z_3$ generalized Kitaev model is defined on the triangular lattice with three types of three-body interactions. Each color represents one type of interaction. } \label{fig:Fig1}
\end{figure}

\section{Model} 
In order to understand the building blocks of our 2D $Z_3$ clock model,  let us first consider the following generalization of the spin-1/2 algebra (Pauli algebra):
\begin{eqnarray}\label{eq:alg-2}
&&\sigma_{z,i}\sigma_{x,i}=\omega \sigma_{x,i}\sigma_{z,i},\quad \sigma_{x,i}\sigma_{y,i}=\omega \sigma_{y,i}\sigma_{x,i},\cr
&& \sigma_{y,i}\sigma_{z,i}=\omega \sigma_{z,i}\sigma_{y,i},\quad \omega=\exp\para{2\pi i /3}.
\end{eqnarray}
along with $\sigma_{a,i}^3=1$, $\sigma_{a,i}^{\dagger}=\sigma_{a,i}^2$, and $\sigma_{x,i}\sigma_{y,i}\sigma_{z,i}=1$ constraints where $a=x,y,z$. In this paper we consider three dimensional irreducible representation of the above algebra (see Appendix A for more detail).

Next, consider the triangular lattice with three sites in the unit cell shown in Fig. \ref{fig:Fig1}. We color the triangular lattice with three different colors: red, green, and blue. In the first model that we consider each color represents a certain three-body interaction among three generalized spins at the corners.
Thus, we define the Hamiltonian as:

\begin{eqnarray}\label{eq:H1}
&& H_{1}=-J_x\sum_{\mbox{R }\Delta\mbox{'s}} T^1_x -J_z\sum_{\mbox{G }\Delta\mbox{'s}}T^1_z-J_y\sum_{\mbox{B }\Delta\mbox{'s}}T^1_y +h.c.\cr
&&\cr
&&T^1_a  \equiv \sigma_{a,i}\sigma_{a,j} \sigma_{a,k},
\end{eqnarray}
where R,G,B stand for red, green, blue.

\section{Slave-parafermion approach} 
Here, we develop a slave-parafermion method to study our model Hamiltonian. Before going into details let us first define the parafermion algebra. $\gamma_{i}$ is called a $Z_n$ parafermion operator when $\gamma_{i}^n=1$, $\gamma_i^\dag=\gamma_i^{n-1}$, and $\gamma_i \gamma_j = e^{2\pi i/n}\gamma_j \gamma_i$ when $i<j$ for a specified ordering~\cite{fradkin1980,fendley2012}. Every two $Z_n$ parafermion operators define an $n-$dimensional Hilbert space, therefore every single parafermion defines a $\sqrt{n}-$dimensional Hilbert space. Now we consider four flavors of $Z_3$ parafermions with 
\begin{eqnarray}
&&\gamma_{x,i} \gamma_{y,i} = \bar{\omega} \gamma_{y,i}\gamma_{x,i},\quad \gamma_{y,i} \gamma_{z,i} = \bar{\omega} \gamma_{z,i}\gamma_{y,i}\cr
&&\gamma_{z,i} \gamma_{x,i} = \bar{\omega} \gamma_{x,i}\gamma_{z,i},\quad \eta_i \gamma_{a,i} = \bar{\omega}\gamma_{a,i} \eta_i .
\end{eqnarray}
local commutation relations~\cite{fradkin1980,fendley2012}. Using the above relations we can represent the generalized spin operators ,$\sigma_{a,i}$'s, in terms of parafermions:  
\begin{eqnarray}
&&\sigma_{x,i}=\gamma_{x,i}^\dag \eta_i,\quad \sigma_{y,i}=\eta_i^\dag \gamma_{y,i} ,\quad \sigma_{z,i}=\gamma_{z,i}^\dag \eta_i.
\end{eqnarray}
It is easy to verify that the above slave-parafermion representations indeed satisfy the algebra in Eq. \eqref{eq:alg-2}. Observe that the above relations enjoy a $Z_3$ gauge symmetry, namely: $\para{\gamma_{i,a},\eta} \to \omega \para{\gamma_{i,a},\eta_i}$ local transformation leaves $\sigma_a$ invariant. Now note that the Hilbert space associated with the clock operators at site $i$ is three dimensional. On the other hand, the dimension of the Hilbert space associated with four parafermions on site $i$ is nine dimensional. As a result there is a three-fold redundancy, hence we must project out the redundant unphysical states. To this end, we can use the $\sigma_{x,i}\sigma_{y,i}\sigma_{z,i}=1$ relation that leads to the following local constraint on the Hilbert space:
\begin{eqnarray}
&&\para{\gamma_{x,i}^\dag \gamma_{y,i}}\para{ \gamma_{z,i}^\dag \eta_i}=1,
\end{eqnarray}
which reduces the total Hilbert space (per lattice sites) by a factor of three. 
In terms of the parafermions, the interaction terms become:
\begin{eqnarray}
&&J_b\sigma_{b,i}\sigma_{b,j}\sigma_{b,k}+h.c.=J_b\para{\gamma_{b,i}\gamma_{b,j}\gamma_{b,k}} \eta_{i}^\dag\eta_{j}^\dag\eta_{k}^\dag+h.c.
\end{eqnarray}
It is straightforward to verify that $P^{b}_{ijk}\equiv \gamma_{b,i}\gamma_{b,j}\gamma_{b,k}=\para{\gamma_{b,i}\gamma_{b,j}^\dag} \para{\gamma_{b,j}^\dag\gamma_{b,k}}$ operators commute with the Hamiltonian as well as among themselves for all $b=x,y,z$ and $i,j,k$'s that form a colored triangle. Consequently, $P_{ijk}^{b}$'s are constants of motion and can be replaced by their expectation values. $P_{i,j,k}^{b}$ takes $Z_3$ values because it cubes to one. Assuming the lowest energy corresponds to uniform value of $P_{ijk}^{b}$'s, we obtain the following low energy effective description of the generalized Kitaev model (GKM) on the triangular lattice:
\begin{align}\label{eq:PFH}
-\sum_{\mbox{R}~\Delta\mbox{'s}} {J}_{x} {\eta_{i}\eta_{j} \eta_{k}}-\sum_{\mbox{G}~\Delta\mbox{'s}} {J}_{z}  {\eta_{i}\eta_{j} \eta_{k}}-\sum_{\mbox{B}~\Delta\mbox{'s}} {J}_{y}  {\eta_{i}\eta_{j} \eta_{k}}+h.c.
\end{align}
The above Hamiltonian suggests that the effective degree of freedom at low energies is described by a single parafermion per site, i.e. there are $3^{N_s/2}$ total degrees of freedom. We can also reach this conclusion by finding the number of conserved $Z_3$ quantities, namely Wilson loop operators. In Appendix B we show that there exist $N_s/2$ commuting distinct Wilson loop operators signaling that half of the degrees of freedom of our 2D clock model are frozen at low energies. 

In order to understand the fate of the above coupled parafermion system we first consider $SU(2)_4$ topological field theory (TFT) that contains five primary fields: $\Phi^l_0$ with $j=l/2$ spin, where $l=0,..,4$~\cite{nayak2008}. Next, we condense the spin-2 field ($\Phi^4_0$) of the TFT~\cite{bais2009}. Doing so, the spin-1/2 ($\Phi^1_0$) and spin-3/2 ($\Phi^3_0=\Phi^1_0 \times \Phi^4_0$) non-Abelian operators become identified and {\em confined}. So $\Phi^1_0\sim \Phi^3_0 \equiv \tau$, where $\tau$ will be referred to as the twist operator. Furthermore, $\Phi^2_0$ branches into $X$, and $Y$ Abelian operators with $X \times X= Y$ and $X \times Y =\mathbb{I}$ fusion rules. Twist operator satisfies $\tau \times \tau = \mathbb{I}+X+Y$ fusion rule and have $d=\sqrt{3}$ quantum dimension accordingly. Therefore, condensing $\Phi^4_0$ field of the $SU(2)_4$ results in $\mathbb{I},X,$ and $Y$ deconfined Abelian and $\tau$ confined non-Abelian excitations~\cite{bais2009}. It can be shown that {\em to each twist operator, $\tau$, a single $Z_3$ parafermion zero mode is attached}~\cite{vaezi2013,barkeshli2013genon}. Thus, we can view the 2D array of parafermion in Eq. \eqref{eq:PFH} as a triangular array of twist fields.

Now let us consider $-\eta_i \eta_j \eta_k=-\para{\eta_i \eta_j^\dag }{\eta_j^\dag \eta_k}$ term in the low energy Hamiltonian, Eq. \eqref{eq:PFH}. A simple analysis shows that these terms favor spin-0 (i.e. $\mathbb{I}$ operator) fusion channel in the fusion of every two neighboring twist fields. 
Hence, the parafermion coupling term, Eq. \eqref{eq:PFH}, can be viewed as a projector onto the spin-0 fusion channel of the $\Phi^1_0 \times \Phi^1_0$ fusion in the $SU(2)_4$ theory (or for $\tau \times \tau$ after $\Phi^4_0$ condensation). In Refs. \cite{ludwig2011,feiguin2007}, the effect of these projectors has been studied and authors have shown that the many body collective state is described by a topological phase with $\frac{SU(2)_3 \times SU(2)_1 }{SU(2)_4}$ edge state with {\em parent $SU(2)_4$ state} and $SU(2)_3 \times SU(2)_1$ with {\em vacuum}. Moreover, authors of Ref. \cite{mong2013} have shown that if we condense $\Phi^4_0$ in the parent $SU(2)_4$ state, the resulting many body state of coupled parafermions will be the Fibonacci phase whose only nontrivial and deconfined excitation is the Fibonacci anyon, $\epsilon$. Thus, we conjecture that the ground-state of the above coupled parafermion system is described by the Fibonacci theory~\cite{trebst2009}. Fibonacci anyons are excitations with $\epsilon \times \epsilon = 1+\epsilon$ fusion rule, $d_F=\para{1+\sqrt{5}}/2\simeq 1.617$ quantum dimension, and $s=2/5$ topological spin. The TFT of the Fibonacci phase is described by an $SU(2)_3 \times SU(2)_1$ Chern-Simons gauge theory and it chiral edge by a $Z_3 \times U(1)_6 \times U(1)_2$ conformal field theory (CFT) with $c=14/5$ central charge, where $Z_3$ stands for the Zamolodchikov-Fateev $Z_3$ parafermion CFT~\cite{zamolodchikov1985}.

\section{Slave fermion approach}
 Here we utilize a different approach, slave-fermion method, to study our GKM. This framework is shown to be quite useful for Kitaev's original model~\cite{burnell2011b}. Since the Hilbert space associated with $Z_3$ clock operators at site $i$ is three dimensional, we can represent them in terms of three flavors of fermions, namely 
\begin{eqnarray}\label{eq:slave-1}
&& \sigma_{i,z}~=\omega^{2}f_{3,i}^\dag f_{3,i}+~\omega f_{2,i}^\dag f_{2,i}~+~~f_{1,i}^\dag f_{1,i},\cr
&&\sigma_{i,x}~=~~f_{1,i}^\dag f_{3,i}~+~~f_{3,i}^\dag f_{2,i} ~~+~~f_{2,i}^\dag f_{1,i},\cr
&&\sigma_{i,y}~=~~f_{3,i}^\dag f_{1,i}~+~\omega f_{2,i}^\dag f_{3,i}~+\omega^2 f_{1,i}^\dag f_{2,i}.
\end{eqnarray}
along with
\begin{eqnarray}\label{eq:slave-2}
&& f_{3,i}^\dag f_{3,i}+f_{2,i}^\dag f_{2,i} + f_{1,i}^\dag f_{1,i}=1,
\end{eqnarray}
constraint that projects states into the physical Hilbert space. 
Note that the Eqs. \eqref{eq:slave-1} and \eqref{eq:slave-2} are invariant under the following $U(1)$ gauge transformation: $f_{n,i} \to e^{i \alpha_i}f_{n,i}$. Furthermore, the model Hamiltonian in Eq. \eqref{eq:H1} is symmetric under the $\para{f_{1},f_{2},f_{3}}\to \para{f_{2}, f_{3}, f_{1}}$ $Z_3$ exchange.

\section{Mean field treatment of the slave fermions} 
Using the slave fermion representation of the clock operators we can easily rewrite the 2D clock Hamiltonian in terms of $f_{n,i}$ fermions. To this end, first note that $\sigma_{a,i}\sigma_{a,j}\sigma_{a,k}=\sigma^\dag_{a,i}\sigma_{a,j}\sigma^\dag_{a,i}\sigma_{a,k}$. Moreover:
\begin{eqnarray}\label{eq:slave-3}
&& \sigma^\dag_{z,i}\sigma_{z,j}=-\sum_{n,m}\omega^{m-n} \hat{\chi}^{n,m}_{i,j} \hat{\chi}^{m,n}_{j,i} ,\cr
&& \sigma^\dag_{x,i}\sigma_{x,j}=-\sum_{n,m}\hat{\chi}^{n,m}_{i,j} \hat{\chi}^{m+1\%3,n+1\%3}_{j,i} ,\cr
&& \sigma^\dag_{y,i}\sigma_{y,j}=-\sum_{n,m}\omega^{n-m} \hat{\chi}^{n,m}_{i,j} \hat{\chi}^{m-1\%3,n-1\%3}_{j,i}.
\end{eqnarray}
where $\hat{\chi}^{n,m}_{i,j}  \equiv f_{n,i}^\dag f_{m,j}$ and $\%$ means indices are defined mod 3.
Now, we would like to use the mean-field approximation and replace $\hat{\chi}^{n,m}_{i,j}$  operator with its expectation value until we reach a quadratic Hamiltonian of slave fermions. For simplicity we assume that the mean-field parameters do not break lattice symmetries. We also make no-fermion-pairing and no-flavor-mixing assumptions so $\braket{\hat{\chi}^{n,m}_{i,j}}=\delta_{n,m} \chi_{i-j}$. Thus, every flavor is conserved and we can promote the $U(1) \times Z_3$ symmetry of the slave-fermion representation to $ U(1) \times U(1) \times U(1) $ symmetry each associated with one flavor conservation. 

\begin{figure}
\centering
{\includegraphics[width=0.5\columnwidth]{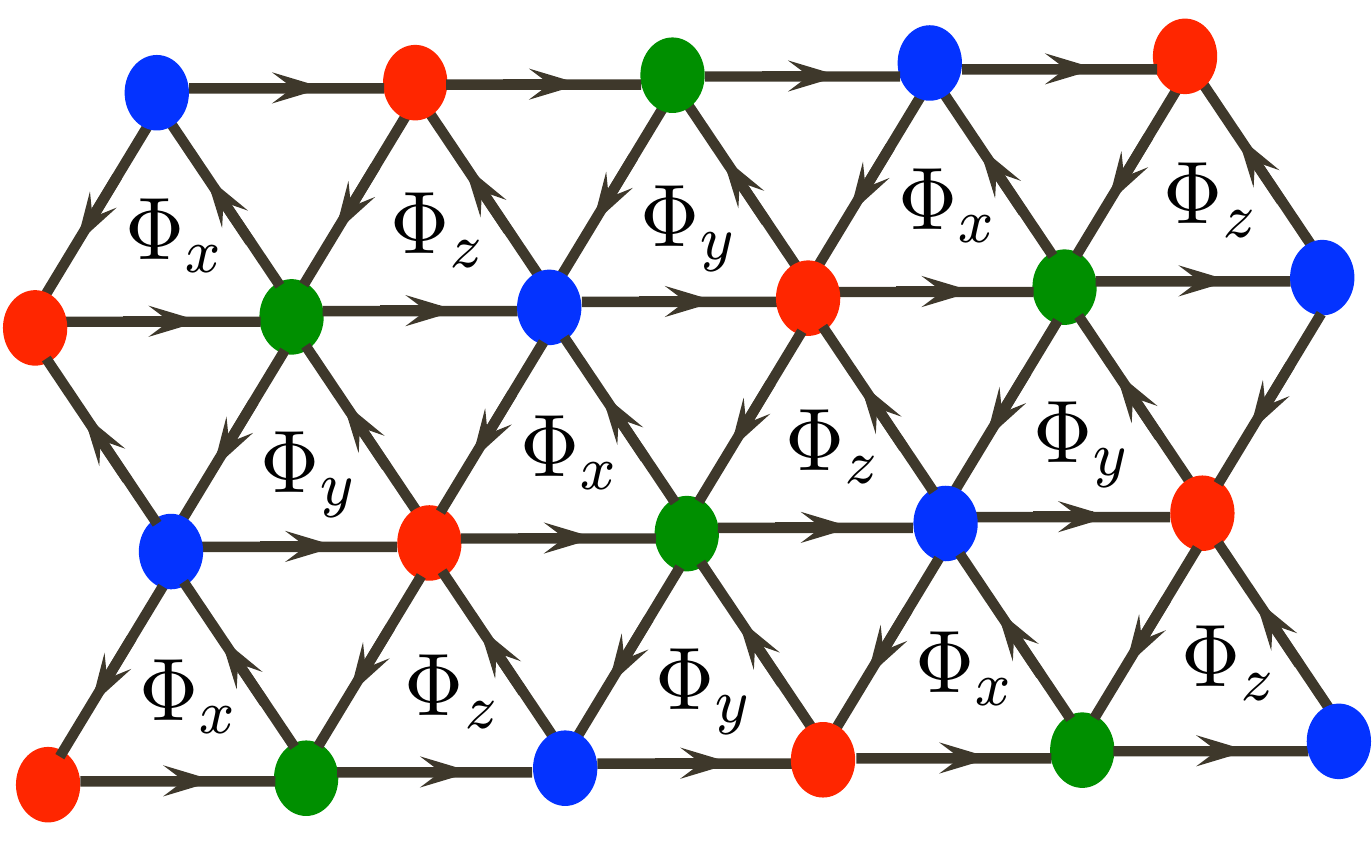}}
\caption{Schematic representation of the no-pairing no-flavor mixing mean-field ansatz. We assume that the mean-field parameters are translation invariant, and do not depend on the flavor. All the mean-field parameters at the edges of a colored triangular have the same value. The flux enclosed by color $a$ triangles, $\Phi_a$, is defined through $\chi_{a}^3=\abs{\chi_a}^3 e^{i\Phi_a}$ identity. We consider three different values for the mean-field parameters and their associated fluxes based on the color.} \label{fig:Fig2}
\end{figure}

\begin{figure}
\centering
{\includegraphics[width=1.0\columnwidth]{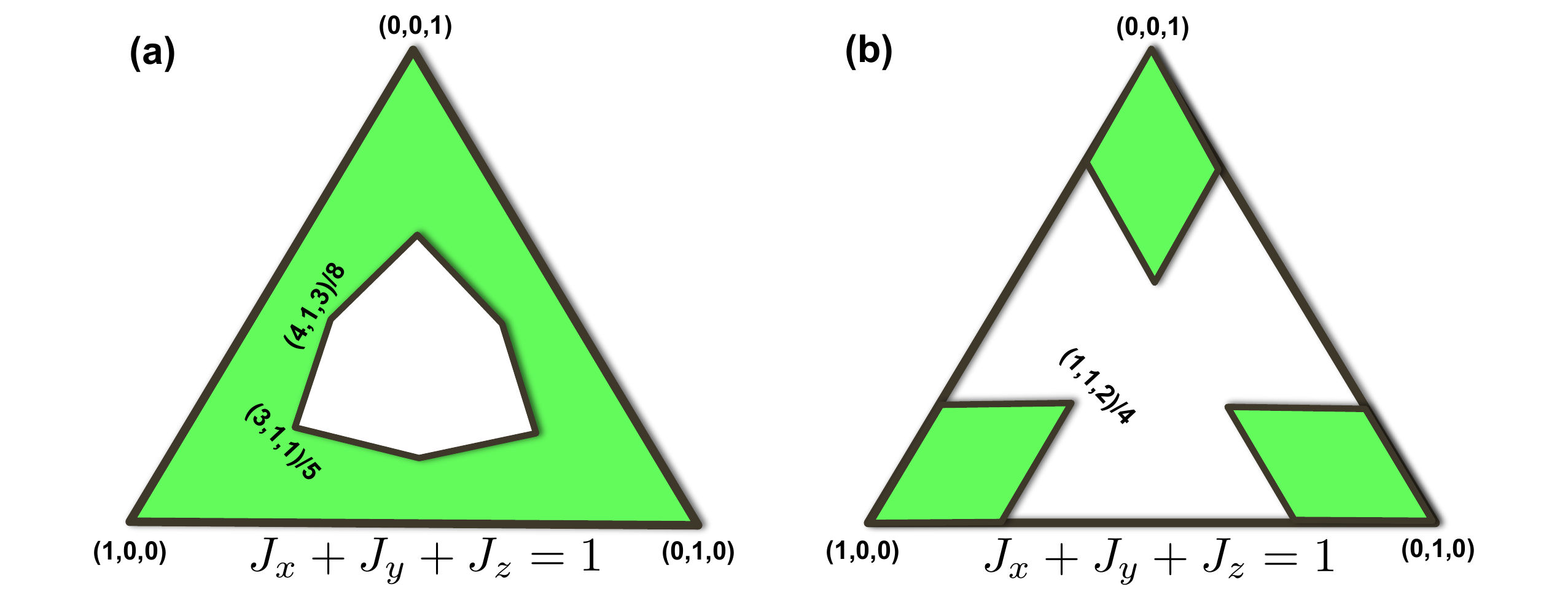}}
\caption{Mean-field phase diagram of the two models discussed in the paper. (a) MF phase diagram of Eq. \eqref{eq:H1}. The equilateral triangle is defined on the $J_x+J_y+J_z=1$ plane and $J_a\geq 0$. White regions denote the topological mean-field solutions, i.e. those areas with $\abs{C}=1$. (b) MF phase diagram of Eq. \eqref{eq:H2}.} \label{fig:Fig3}
\end{figure}

Now, we solve the mean field equations self-consistently to obtain the mean field parameters. From Eq. \eqref{eq:slave-3} and Fig. \ref{fig:Fig2}, this mean-field ansatz results in the following mean-filed Hamiltonian:  
\begin{eqnarray}
 &&H^1_{MF}= \sum_{k} \psi^\dag_{n,k} h\para{k_1,k_2} \psi_{n,k}.
\end{eqnarray}
where  $J^*_{a}=3J_{a}\abs{\chi_{a}}^2$, $\psi_{n,k}^{\rm T}= \para{f_{z,k},f_{y,k},f_{x,k}}$, and 
\begin{eqnarray}
&& h_{1,2} = J^*_x \chi_x e^{-ik_1}+J^*_y \chi_y e^{-ik_2}+J^*_z \chi_z \cr 
&& h_{2,3}=  J^*_x \chi_x e^{-ik_2}+J^*_y \chi_y+J^*_z \chi_z e^{-ik_1}\cr
&& h_{3,1}=  J^*_x \chi_x e^{i\para{k_1+k_2}}+ J^*_y \chi_y e^{ik_2}+J^*_z \chi_z e^{-ik_1}.
\end{eqnarray}
 Using the above Hamiltonian, the mean-field parameters can be solved self-consistency through $\chi^{n,m}_{i,j}  \equiv \langle f_{n,i}^\dag f_{m,j}\rangle$ relations. 
Next, we compute the Chern number~\cite{qi2010RMP} associated with each flavor's band-structure (see Appendix C material for more detail). To this end, first recall that the local constraint on the Hilbert space in Eq. \eqref{eq:slave-2} requires every site to contain one slave fermion. Therefore, due to the symmetry of the mean field ansatz, the average number of a certain fermion flavor per unit cell is one and the lowest energy band is fully occupied for every flavor of fermions. If the lowest band is separated from higher energy bands by a finite energy gap, then we can assign a topological Chern number to it. 
The mean-field phase diagram is shown in Fig. \ref{fig:Fig3}.

\section{Low energy description: Chern-Simons (CS) gauge theory} 
To implement the local constraint on the Hilbert space in Eq. \eqref{eq:slave-2} it is easier to perform a particle hole transformation on one fermion flavor e.g. $f_3$. Defining $d_{3,i} \equiv f_{3,i}^\dag$ operator, the local constraint becomes: $d_{3,i}^\dag d_{3,i}=f_{2,i}^\dag f_{2,i} + f_{1,i}^\dag f_{1,i}$. Given the fact that the hopping Hamiltonian, $h_{i,j} $, transforms to $-h_{j,i}=-h^{\rm T}$ under the particle hole transformation we can immediately see that the Chern number remains invariant during this transformation. Due to the no-flavor mixing symmetry of the mean-field ansatz we can define three auxiliary $U(1)$ gauge fields, $a_{1,\mu}, a_{2,\mu} $, and $a_{3,\mu}$, that are minimally coupled to $f_1, f_2$, and $d_3$ slave fermions, respectively. These auxiliary gauge fields are related to the current density of the slave fermions of flavor $n$ (if $n=1,2$) or its particle-hole conjugate (if $n=3$) through $J^{\mu}_n= \frac{1}{4\pi}\epsilon^{\mu\nu\lambda} \partial_{\nu} a_{n,\lambda}$ relation~\cite{wen1991prl,wen1999}. Such relations guarantee the flavor conservation symmetry because $\partial_{\mu} J^{\mu}_{n}=0$. Knowing the Chern number, we can integrate the massive $f_1, f_2$ and $d_3$ fermions to achieve the effective low energy description of the system in terms of CS theory in terms of auxiliary gauge fields~\cite{wen1991prl,wen1999}. Doing so, we obtain: 
\begin{eqnarray}\label{eq:CS-1}
&& \mathcal{L}=\sum_{n}\mathcal{L}_n=\sum_{n} \frac{C}{4\pi}\epsilon^{\mu\nu\lambda} a_{n,\mu} \partial_{\nu} a_{n,\lambda} .
\end{eqnarray}  
  
The next important step is to enforce the local constraint in Eq. \eqref{eq:slave-2} within the CS gauge theory. In terms of current densities the local constraint becomes $J^{\mu}_3=J^{\mu}_1+J^{\mu}_2$. This condition on the Hilbert space can be translated in auxiliary gauge fields language. We can fix the gauge such that:
\begin{eqnarray}
&& a_{3,\mu}=a_{1,\mu}+a_{2,\mu}.
\end{eqnarray}
Therefore, $f_{1}$ ($f_{2}$) carries unit charge under $ a_{1,\mu}$ ($ a_{2,\mu}$) auxiliary gauge field and is neutral under $ a_{2,\mu}$ ($ a_{1,\mu}$). On the other hand, $f_{3}=d_{3}^\dag$ carries a negative unit charge under both $ a_{1,\mu}$ and $a_{2,\mu}$ auxiliary gauge fields. Plugging the above relation for $a_{3,\mu}$ in Eq. \eqref{eq:CS-1}, the total CS Lagrangian becomes:
\begin{eqnarray}\label{eq:CS-2}
&& \mathcal{L}_{\rm CS} = \frac{K_{n,m}}{4\pi}\epsilon^{\mu\nu\lambda} a_{n,\mu} \partial_{\nu} a_{m,\lambda} ,~~~~K= C\para{  \begin{array}{cc}
    2 & 1 \\ 
    1 & 2 \\ 
  \end{array}}.~~
\end{eqnarray}
The above CS action is exactly identical to that of a $(221)$ bilayer quantum Hall state for $\abs{C}=1$. The $K$ matrix fully determines the topological properties of the ground-states for the Abelian quantum Hall states~\cite{wen1995}. For instance, the topological degeneracy of the ground-state on a genus $g$ manifold is $\abs{K}^g$. Furthermore, the anyon excitations are labeled by integer valued $\vec{l}$ vector, whose self statistics is $\theta_{ll}=\pi {\bf l}^{\rm T}K^{-1}{\bf l}$. The mutual statistics between two different excitations is: $\theta_{ll'}=2\pi {\bf l}^{\rm T}K^{-1}{\bf l'}$. For the above $K$ matrix, besides the trivial excitation, $\vec{l}=(0,0)$, there are two other non-trivial anyon excitations: $\para{l_1,l_2}=(1,0),(1,1)$, both with $\theta=2\pi/3$ self-statistics. 
Furthermore, there are two electron excitations: $\psi_1$ with $\vec{l}=(2,1)$ vector and $\psi_2$ with $\vec{l}=(1,2)$. 
The edge CFT of the $(221)$ state is described by two free bosons $\phi_1$ and $\phi_2$. From the bulk wavefuction-edge CFT duality we can bosonize different fermion flavors after which:
$ f_{1} \sim e^{i\phi_1},~f_{2} \sim e^{i\phi_2},$ and $f_{3} \sim e^{-i\para{\phi_1+\phi_2}}$. Therefore, all slave fermions are fractional excitations with $\theta=2\pi/3$ self-statistics. Similarly, we can find the free boson representation of electron operators and we have: 
\begin{eqnarray}\label{eq:frac-1}
&&\psi_1\sim e^{i\para{2\phi_1+\phi_2}}\sim f_{3}^\dag  f_{1}, ~~~ \psi_2\sim e^{i\para{\phi_1+2\phi_2}}\sim f_{3}^\dag f_{2}. ~~~
\end{eqnarray}

\section{Beyond mean-field result: Fibonacci phase} 
So far, we approximated the GKM in Eq. \eqref{eq:H1} with a quadratic fermion Hamiltonain. Doing so, we achieved the CS low energy effective theory of the model. We showed that for a large part of the phase diagram, the CS theory is identical to that of a $(221)$ bilayer FQH state. Now we would like to study fluctuations beyond the mean field by taking the effect of quartic fermion terms into considerations.  A generic quartic term can be obtained from $\sigma_{a,i}^\dag \sigma_{a,j} \sigma_{a,i}^\dag \sigma_{a,k} \simeq \braket{\sigma_{a,i}^\dag \sigma_{a,k}} \sigma_{a,i}^\dag \sigma_{a,j}$ approximation and using the slave fermion representation for $\sigma_{a,i}^\dag \sigma_{a,j}$ operator. The $\sigma_{z,i}^\dag \sigma_{z,j}$ term generates terms of $f_{n,i}^\dag f_{n,i} f_{m,j}^\dag f_{m,j}$ form which acts like a density-density interaction on the $(221)$ state. On the other hand, the $\sigma_{x,i}^\dag \sigma_{x,j}$ and $\sigma_{y,i}^\dag \sigma_{y,j}$ terms generate $f_{n,i}^\dag f_{n+1,i} f_{m+1,j}^\dag f_{m,j}$ quartic terms. Every resulting term can be easily investigated from the free boson representation of the slave fermion and electron operators in Eq. \eqref{eq:frac-1}. For example, $f_{3,i}^\dag f_{1,i} f_{3,j}^\dag f_{2,j} \sim \psi_{1,i}\psi_{2,j}$ acts like an interlayer pairing for the $(221)$ bilayer state. Therefore, the quartic perturbations around the mean-field solutions can be mapped to the $(221)$ bilayer state with uniform interlayer pairing as well as density-density interaction (of both interlayer and intralayer types). In Refs. \cite{mong2013} and \cite{vaezi2013b}, the problem of perturbing a 2/3 FQH state with interlayer pairing has been studied and the authors show that the non-Abelian Fibonacci phase can emerge for a strong enough pairing. A related study of the $2/3$ FQH perturbed by uniform interlayer tunneling in Ref. \cite{vaezi2014} has also shown the emergence of the Fibonacci phase above some threshold (see Appendix D for detailed analysis). Therefore, we conjecture that the ground-state of the GKM in Eq. \eqref{eq:H1} can belong to the Fibonacci phase. 

\section{Another related model Hamiltonian} 
Here we introduce another $Z_3$ generalizations of the Kitaev's honeycomb model on the triangular lattice:
\begin{eqnarray}\label{eq:H2}
&& H_{2}=J_x\sum_{\mbox{R }\Delta\mbox{'s}} T^2_x + J_z\sum_{\mbox{G }\Delta\mbox{'s}}T^2_z+J_y\sum_{\mbox{B }\Delta\mbox{'s}}T^2_y +h.c.\cr
&&\cr
&&T^2_a  \equiv { \sigma^\dag_{a,i}\sigma_{a,j}+\sigma_{a,j}^\dag \sigma_{a,k}+\sigma_{a,k}^\dag \sigma_{a,i}}.
\end{eqnarray}
Unfortunately the slave-parafermion method does not simplify the above model, though we still can find enough Wilson loop operators. On the other hand, the the slave-fermion framework works equally well for this model and we can obtain the ground-state properties by following similar procedures we performed for the model Hamiltonian in Eq. \eqref{eq:H1}. The results of such a mean-field  analysis is presented in Figs. \ref{fig:Fig3}. Again, the fluctuations above the mean-field solution lead to interlayer pairing in the dual (221) state, hence the Fibonacci phase is a possible ground-state of this Hamiltonian.

\section{Conclusion} 
We presented arguments based on the slave-parafermion and slave-fermion approaches supporting our conjecture that there exists a family of three-state clock Hamiltonians with topological spin liquid ground-state and protected chiral edge states for a wide range of coupling constants.  This fractionalized spin liquid itself can belong to two distinct phases: Abelian state or non-Abelian. The Abelian phase of this topological spin liquid is dual to a $(221)$ bilayer FQH state with similar fractional excitations, while its non-Abelian state is a Fibonacci phase with non-Abelian excitations. These results together with the Kitaev's original honeycomb model suggests a deeper relation between 2D $Z_{n}$ clock models with strong anisotropic interactions and fractional topological superconductors with an edge state containing $Z_{n}$ parafermion CFT.  

Before closing we like to mention that the methods developed in this paper are less useful for the honeycomb model introduced in Ref. \cite{BarkeshliGKM}. For instance, the slave parafermion method is powerful for our model because terms likes
$P^{b}_{ijk}= \gamma_{b,i}\gamma_{b,j}\gamma_{b,k}$
are constants of motion and as a result the remaining degrees of freedom depend on one flavor of parafermions only. However, a simple calculation shows that there are no such conserved quantities in terms of slave parafermions for two-body interacting $Z_3$ clock models. Furthermore, the slave-fermion approach is less straightforward for the honeycomb model. If we assume the mean-field Hamiltonian of spinons do not break lattice symmetries, then the unit cell would contain two sites and as a result the filling of the lowest band for each flavor of slave fermions would be 2/3. Hence, the lowest band is partially occupied for this ansatz and we cannot assign Chern number to it. However, we can assume a mean-field ansatz where a lattice symmetry e.g. translation symmetry is broken to enlarge the unit cell such that an integer number of bands become filled. This type of mean-field state can be achieved by assuming a staggered flux mean-field Hamiltonian for example. The lattice symmetry can be restored after the projection to the physical Hilbert space, therefore the symmetries are realized projectively. It is interesting to study whether these types of mean-field ansatz are energetically favorable and also what kind of topological field theory would describe the resulting projected state.

\section{Acknowledgments}
This work was inspired by Xiao-Liang Qi's talk in Simons Center \footnote{Xiao-Liang Qi, presentation on the workshop Topological Phases of Matter Workshop at Simons Center,
Stony Brook University, June 13th, 2013. http://scgp.stonybrook.edu/archives/3464}. We gratefully acknowledge useful discussions with Michael Lawler, Shivam Ghosh, Eun-Ah Kim, Paul Fendley and Jason Alicea. This work was supported by the Bethe postdoctoral fellowship.  



\section{Appendix A. Generalized spin operators}
Original Kitaev's honeycomb model can be generalized in different way using the spin-1/2 operators~(see for example Refs. \cite{Yao2007,Bombin2009,Mandal2009,Wu2009}). In this paper, however, we would like to make a different generalization by introducing operators that satisfy a different commutation algebra than spin 1/2's. In this appendix we give the matrix representation of the $Z_3$ algebra utilized in this paper. To that end, let us first consider the following two matrices: $\sigma\equiv\left(\begin{array}{ccc}
    1 & 0 & 0 \\ 
    0 & \omega & 0 \\ 
    0 & 0 & \omega^2 \\ 
  \end{array}\right)$ and $\tau\equiv\left(\begin{array}{ccc}
    0 & 0 & 1 \\ 
    1 & 0 & 0 \\ 
    0 & 1 & 0 \\ 
  \end{array}\right)$ where $\omega=e^{2\pi i/3}$. These matrices form the following $Z_3$ commutation algebra:
\begin{eqnarray}
&&\sigma \tau=\omega \tau \sigma,\quad \sigma^3=\tau^3=1,\cr
&&\sigma^\dag=\sigma^2,\quad \tau^\dag=\tau^2.
\end{eqnarray}

The generalized spin operators are defined in terms of the above $Z_3$ clock operators as follows:
\begin{eqnarray}
&&\sigma_{x,i}\equiv \tau_i,\quad \sigma_{z,i}\equiv \sigma_i,\quad \sigma_{y,i}\equiv \tau_{i}^\dag \sigma_{i}^\dag.
\end{eqnarray}

The above three operators satisfy the following algebra:
\begin{eqnarray}\label{eq:App:alg-2}
&&\sigma_{z,i}\sigma_{x,i}=\omega \sigma_{x,i}\sigma_{z,i},\cr
&& \sigma_{x,i}\sigma_{y,i}=\omega \sigma_{y,i}\sigma_{x,i},\cr
&&\sigma_{y,i}\sigma_{z,i}=\omega \sigma_{z,i}\sigma_{y,i}.
\end{eqnarray}
which resembles the spin-1/2 algebra except for the $-1 \to \omega$ substation.

\section{Appendix B. Conserved Wilson loop operators}
Now we show that there are $N_{s}/2$ conserved Wilson loop operators, where $N_s$ is the number of sites. Since, each Wilson loop reduces the dimension of low energy subspace by a factor of three, the total degree of freedom in the low energy subspace is $3^{N_s/2}$ which clearly points towards a single parafermion degrees of freedom per lattice site.

\begin{figure}
\centering
{\includegraphics[width=0.8\columnwidth]{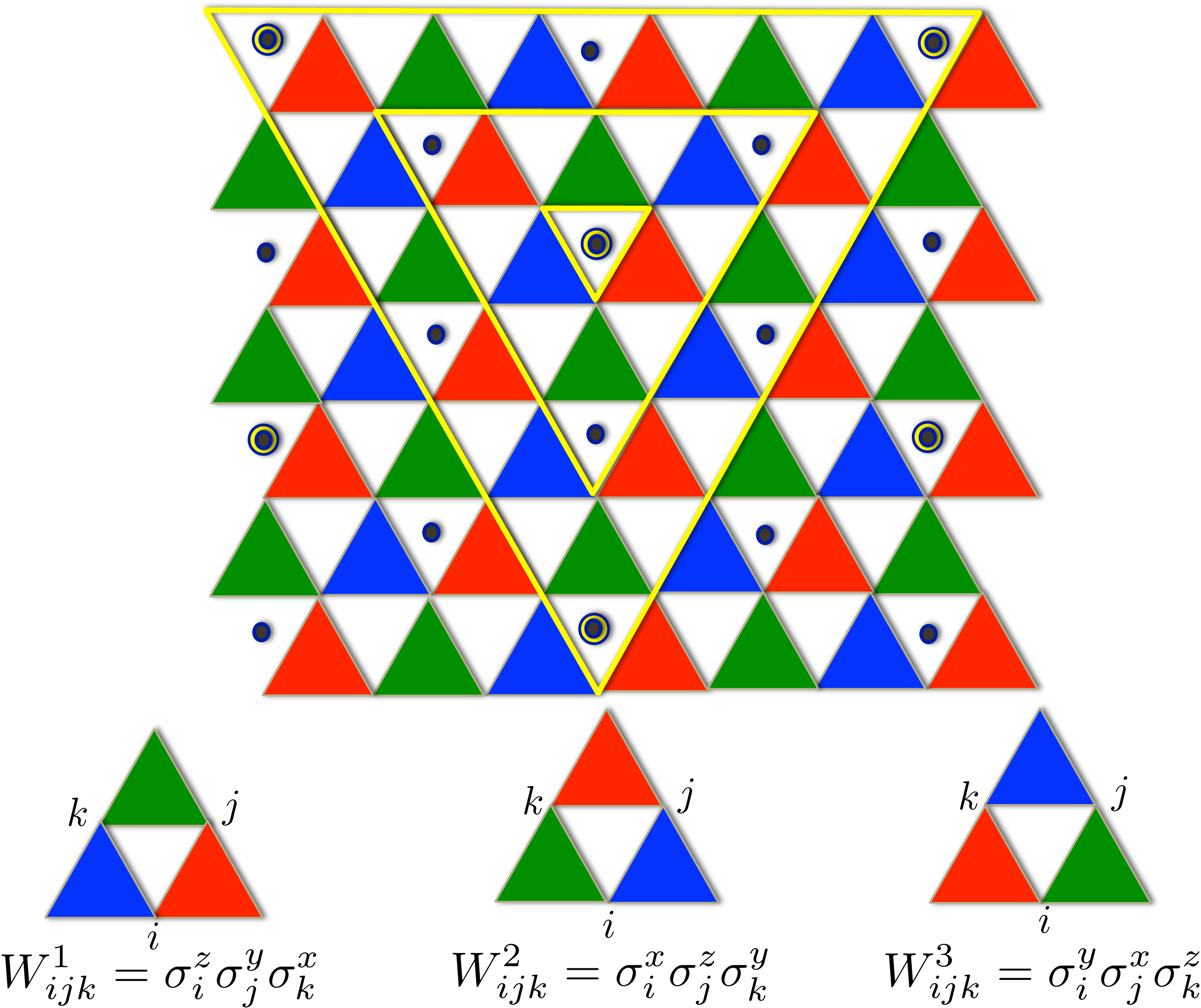}}
\caption{Conserved Wilson loop operators.-- There are three types of white triangles in the lattice, each represented by a different Wilson loop. However, every three triangles that meet at a point are non-commuting and we must pick one of them only. As a results, every dotted white triangle (including those with bigger black circles) represents an allowed Wilson loop operator. They account for $N_{s}/3$ conserved quantities. The triangles with bigger dots host two larger wilson loops defined by yellow lines in addition to the small triangles. These larger Wilson loops are defined by the product of every white triangle's Wilson loop that they enclose. Thus we obtain $N_{s}/12+N_{s}/12=N_{s}/6$ additional conserved quantities. Therefore, altogether, we find $N_{s}/2$ $Z_3$ conserved quantities.}  \label{fig:Fig4}
\end{figure}

Let us consider uncolored (white) triangles. There are three types of them based on the color of the triangle above their top edge. For example, let us consider the white triangle whose top edge is next to a green triangle (type I). Next, we assign the following Wilson loop operator to it: $W^1_{ijk}=\sigma^z_i\sigma^y_j\sigma^x_k$. It is straight forward to verify that this loop operator commutes with all terms in the Hamiltonian. So it can be a constant of motion. Similarly, we can assign a different Wilson loop operator to a white triangle whose top edge is next to a green triangle as follows: $W^{2}_{ijk}=\sigma^x_i\sigma^z_j\sigma^y_k$. Although this loop operator also commutes with the Hamiltonian, it does not commute with $W^1_{ijk}$, hence we have to choose either of them, not both as a constant of motion. We could also consider $W^{3}_{ijk}=\sigma^y_i\sigma^x_j\sigma^z_k$ operator for type III white triangles, but it does not commute with Wilson loops type I neither type II though it commutes with the Hamiltonian. Therefore, we can take only type to be a constant of motion, say $W^1_{ijk}$. This immediately leaves us with $N_{1}=N_{s}/3$ small Wilson loops marked with blue dots in Fig \ref{fig:Fig4}. Although any two corner-sharing Wilson loop operators (which are of different types) have non-trivial commutation relations and thus cannot be simultaneously conserved, we can consider larger loops that are made of several small Wilson loops and are defined as their products can be shown to commute with small $W^1_{ijk}$ operator. A simple analysis shows that there are two kinds of larger Wilson loops, each with $N_s/12$ abundance that commute with the Hamiltonian, themselves and other Wilson loop operators. Therefore, in total we can find $N_{s}/2$ commuting Wilson loops that can are constants of motion and therefore reduce the dimension of the Hilbert space by a factor of $1/3^{N_s/2}$ (Wilson loops take $Z_3$ values as they are defined through $Z_3$ clock operators). As a result, we are left with $\sqrt{3}$ degrees of freedom per lattice site at low energies that implies the existence of a single $Z_3$ parafermion local degree of freedom coupled to a $Z_3$ gauge field due to the condensation of Wilson loops.

\section{Appendix C. Computing Chern number of the slave fermions}

Recall that we had to impose the following constraint on the number of slave fermions at every lattice site:
\begin{align}
f_{1,i}^\dag f_{1,i}+f_{2,i}^\dag f_{2,i}+f_{3,i}^\dag f_{3,i}=1.
\end{align}
Assuming a translational symmetric mean-field ansatz we obtain $\langle f_{1,i}^\dag f_{1,i} \rangle=\langle f_{2,i}^\dag f_{2,i} \rangle=\langle f_{3,i}^\dag f_{3,i} \rangle=1/3$. Furthermore, the unit cell on our decorated triangular lattice contains three sites. Therefore, the average number of slave fermions of a certain flavor per unit cell is 1, hence the lowest band first Brillouin zone is completely filled. Due to the number of three sites in the unit cell, we obtain three energy bands. Let us denote the lowest energy band of the mean-field Hamiltonian associated with flavor $a$ fermion by $\ket{{\bf k},1}_{a}$, and similarly the two higher energy bands by $\ket{{\bf k},2}_a$ and $\ket{{\bf k},3}_a$. Based on what we just argued the lowest band is fully occupied and if it is well separated by a nonzero gap from $\ket{{\bf k},2}_a$ band, we can index it with the Chern number. This topological index can be obtained through the following steps:

{\em \quad a. Find the Berry connection defined as:  $\mathcal{A}^{a}_{\mu}\para{\bf k}=i\bra{k,1}_a\frac{\partial}{\partial k_{\mu}}\ket{k,1}_a$}

{\em \quad b. Compute the Berry curvature through:  $\mathcal{F}^{a}_{k_xk_y}\para{\bf k}=\frac{\partial \mathcal{A}^a_{k_y}}{\partial k_{x}}-\frac{\partial \mathcal{A}^a_{k_x}}{\partial k_{y}}$ relation}

{\em \quad c. Obtain the Berry phase by integrating the Berry curvature over the first Brillouin zone. } 

{\em \quad d. Chern number is related to the Berry phase in the following way:}

\begin{align}
C^{a}=\frac{\theta^a_{\rm B}}{2\pi}=\frac{1}{2\pi}\int dk_x dk_y \mathcal{F}^a_{k_xk_y}.
\end{align}

In the $Z_3$ symmetric state, all flavors would have the same Chern number. As an example, let us apply the following rules to a two band model and compute the Chern number associated with its valence band. To this end consider the following Bloch Hamiltonian:

\begin{eqnarray}
H_{k}=-\vec{\tau}.\vec{d}_{k}
\end{eqnarray}
where $\sigma_i$ are Pauli matrices. The negative eigenvalue of $H_{k}$ is $E_{k,-}=-\abs{d_{k}}$ and its associated eigenstate is: 

\begin{eqnarray}
\ket{k,-}=\para{  \begin{array}{c}
    e^{-i\phi_{k}/2}\cos\para{\theta_k/2} \\ 
    e^{i\phi_{k}/2}\sin\para{\theta_k/2} \\ 
  \end{array}}
\end{eqnarray}
where $\cos\para{\theta_k}=\hat{d}_k.\hat{z}$ and $\tan\para{\phi_k}=\frac{\hat{d}_k.\hat{y}}{\hat{d}_k.\hat{x}}$, where $\hat{d}_{k}=\vec{k}/\abs{k}$. According to the definition of the Berry connection it would be:

\begin{eqnarray}
&&{\mathcal{A}}_{k_i}\para{k}=\frac{1}{2}\cos\para{\theta_k}{\partial}_{k_i} \phi_{k},
\end{eqnarray}
which results in the $\mathcal{F}_{k_xk_y}\para{\vec{k}}=\frac{1}{2}\hat{z}.\para{\vec{\nabla}\cos\para{\theta_k}\times \vec{\nabla} \phi_{k}}$ expression for the Berry curvature. We can rewrite this latter relation in terms of the $\hat{d}_{k}$ vector after which we achieve the following celebrated formula for the Berry curvature of a two band Hamiltonian: 
\begin{eqnarray}
&&\mathcal{F}_{k_xk_y}\para{\vec{k}}= \frac{1}{2}\hat{d}_{k}.\para{\partial_{k_{x}}\hat{d}_{k}\times \partial_{k_y}\hat{d}_{k}}.
\end{eqnarray}
Accordingly, the Chern number assigned to the lower band is $C=\frac{1}{4\pi}\int dk_x dk_y \hat{d}_{k}.\para{\partial_{k_{x}}\hat{d}_{k}\times \partial_{k_y}\hat{d}_{k}}$ which is the winding number associated with the mapping of the 2D torus (the 1st Brillouin zone) to the unit sphere spanned by $\hat{d}_k.$

\section{Appendix D. Phase diagram of the (221) FQH state with uniform inter-layer electron pairing}

In this section, we briefly discuss the fate of the (221) Halperin state perturbed by uniform interlayer electron pairing. In reference \cite{vaezi2014}, we have studied a related problem: $(nnl)$ FQH in the presence of interlayer tunneling. Using a variety of distinct approaches, we obtained a phase transition to a non-Abelian state with $U(1)_{2(n+l)}\times SU(2)_{n-l}$ edge CFT. On the other hand, in the same paper, we showed that $(nnl)$ FQH with interlayer pairing problem can be mapped to $(nn,-l)$ FQH perturbed by interlayer tunneling that for example can be justified by making particle hole-transformation on one layer. Therefore, $(221)$+pairing is dual to $(22,-1)$+tunneling problem whose fate is the Fibonacci phase with $U(1)_{2}\times SU(2)_3$ edge CFT. Accordingly, we conjecture that taking the effect of interlayer pairing on $(221)$ Halperin state into consideration yields a phase transition to the Fibonacci phase. For completeness, in the remainder of this section we present the coupled wire construction approach discussed in Refs. \cite{mong2013,vaezi2013b,teo2011,Klinovaja2013} that can help us understand the phase transition to the Fibonacci phase better.

\begin{figure}
\includegraphics[width=.5\textwidth]{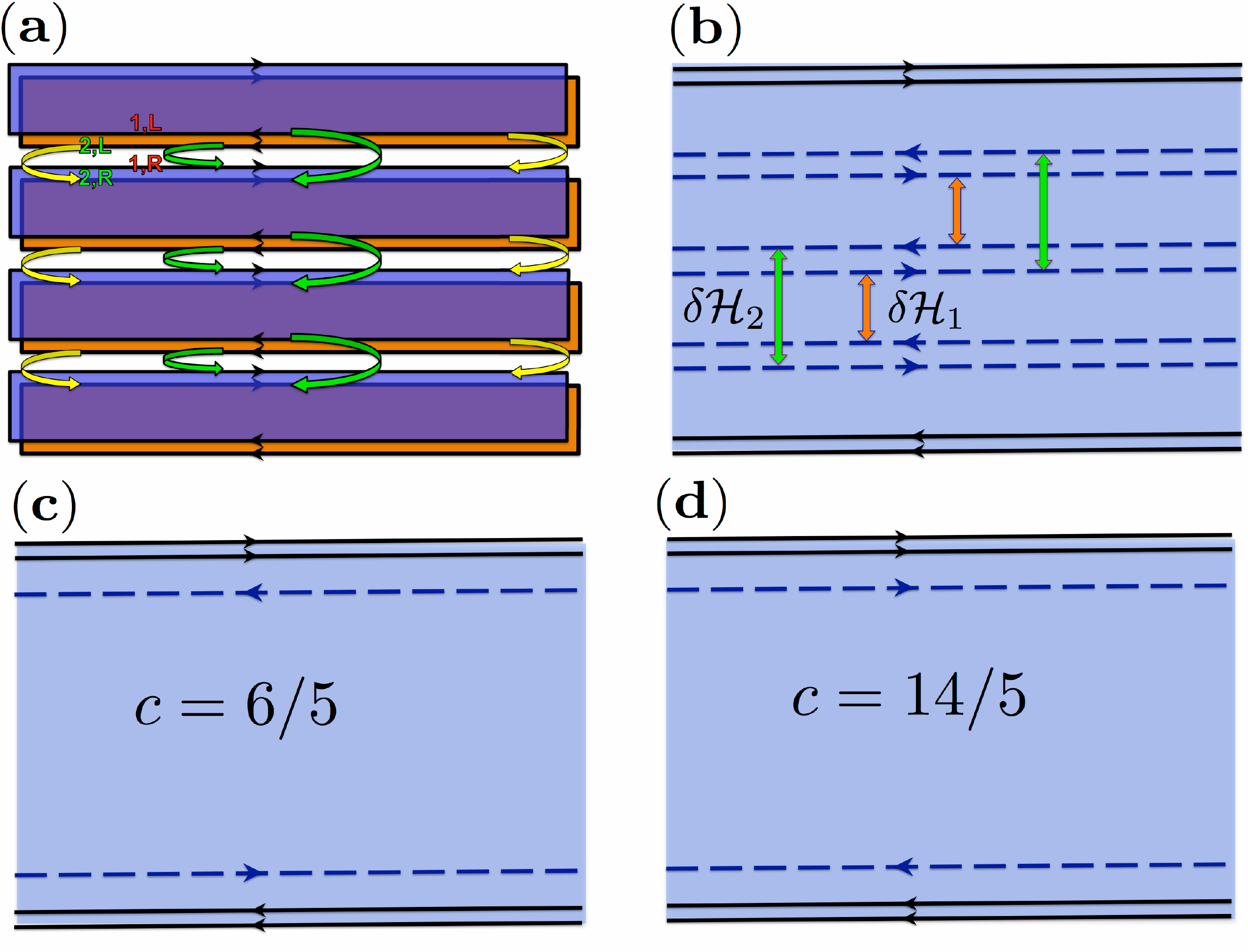}
 \caption{Coupled wire construction in $(221)$ state. (a) The (yellow) green arrows 
represent the (intra-wire electron-backscattering) inter-wire electron pairing. Note that the outermost free boson modes 
do not couple to other modes, and remain gapless. (b) By tuning the 
strengths of the two different types of backscattering terms, we obtain counterpropagating 
gapless $Z_3$ parafermion modes, shown in dotted lines. We have drawn the right and 
left moving parafermion modes to be spatially separated, although their spatial profile
may be more complicated. The parafermion chains can in principle couple in two different 
ways, shown pictorially by the blue and orange arrows. These couplings may require strong electron tunneling,
in addition to allowed quasiparticle tunnelings, and can gap the counterpropagating modes. (c) When the strength of orange 
type $(\delta \mathcal{H}_1)$ inter-wire coupling dominates, the topmost anti-chiral parafermion 
mode remains gapless. This case corresponds to $c=6/5$ total central charge of the chiral edge 
CFT. (d) When the strength of green type inter-wire coupling $(\delta \mathcal{H}_2)$ dominates, 
the topmost chiral parafermion remains gapless. This case corresponds to $c=14/5$ total 
central charge of the chiral edge CFT.}
\end{figure}\label{fig:Fig5}

\noindent{\bf Coupled wire approach to (221) state with uniform inter-layer electron pairing}
Let us consider two adjacent (221) bilayer FQH bars. At their interface, we have two left-moving chiral modes
from the upper bar, and two right-moving anti-chiral modes on the lower bar (see Fig. \ref{fig:Fig4}). In our notation $\phi_{I,R}$, for $I = 1,2$ is the right-moving boson on the 
$I$th layer, while $\phi_{I,L}$ is the left-moving boson from the $I$th layer. It is convenient to define the linear combinations:
\begin{align}
\phi_{cR} &= \sqrt{\frac{3}{2}} (\phi_{1,R} + \phi_{2,R}),
\nonumber \\
\phi_{sR} &= \frac{1}{\sqrt{2}} (\phi_{1,R} - \phi_{2,R}),
\end{align}
and similarly for the left moving modes. These describe the charged and neutral modes, respectively. 
Using the $K$ matrix, it is easy to verify  that the new bosonic fields are compactified on circles of radius $R_c = \sqrt{6}$ and $R_s=\sqrt{2}$: $\phi_{c/s,R} \sim \phi_{c/s,R} + 2\pi R_{c/s}$, and similarly
for $\phi_{c/s,L}$. The electron destruction operator on each of the gapless modes is:
\begin{eqnarray}
&&\Psi_{1,R} \propto e^{2i\phi_{1,R}+\phi_{2,R}} \equiv e^{ i{\sqrt{\frac{3}{2}}}\phi_{c,R}+i{\sqrt{\frac{1}{2}}}\phi_{s,R}} \cr
&&\Psi_{2,R} \propto e^{\phi_{1,R}+2i\phi_{2,R}} \equiv e^{ i{\sqrt{\frac{3}{2}}}\phi_{c,R}-i{\sqrt{\frac{1}{2}}}\phi_{s,R}},
\end{eqnarray}
and similarly for the left-moving modes. For simplicity we assign $\psi_{1}$ to the top and $\psi_{2}$ to the bottom layer. The Hamiltonian that describes these four gapless modes in the absence of any perturbation is:
\begin{eqnarray}
\mathcal{H}_{0} =\sum_{\tau=c,s} \frac{1}{4\pi}\int dx \para{\para{\partial_x \varphi_{\tau}}^2+\para{\partial_x \theta_{\tau}}^2}.
\end{eqnarray}
where $\varphi_{c/s}=\frac{\phi_{c/s,R}+\phi_{c/s,L}}{\sqrt{2}}$, $\theta_{c/s}=\frac{\phi_{c/s,R}-\phi_{c/s,L}}{\sqrt{2}}$ 
are conjugate bosonic variables. Next, we consider the following perturbations, corresponding to intralayer electron backscattering and interlayer electron pairing between counter-propagating gapless modes:
\begin{eqnarray}
\delta \mathcal{H}_{\parallel} = && -t_{\parallel}\para{\Psi_{1,R}^\dag \Psi_{1,L}+\Psi_{2,R}^\dag \Psi_{2,L} }+H.c.\cr
\delta \mathcal{H}_{\perp} =&&-\Delta_{\perp}\para{ \Psi_{1,R}^\dag \Psi^\dag_{2,L}+\Psi_{2,R}^\dag \Psi^\dag_{1,L}}+H.c.
\end{eqnarray}
We can translate the above perturbations in terms of bosonic fields, after which we have:
\begin{eqnarray}
\delta \mathcal{H} = -4 \cos\para{\theta_{s}}\para{\Delta_{\perp}\cos\para{\sqrt{3} \varphi_{c}}+t_{\parallel}\cos\para{\sqrt{3}\theta_{c}}}.
\end{eqnarray}
In the absence of pairing i.e. for $\Delta_{\perp}=0$, every two counter-propagating modes that are coupled with backscattering become gapped. The generated gap that also defines the bulk gap of the fractional quantum Hall state is therefore of order $t_{\parallel}$. However, the pairing term competes with the backscattering term and decreases the bulk gap. At some point the bulk gap closes and the system undergoes a phase transition. For larger values of $\Delta_{\perp}$ the bulk is again non-zero, but the topological oder of the system is changed. The above effective Hamiltonian can be greatly simplified if we drop the $\cos\para{\theta_s}$ term. In fact this can be done formally due to the following argument.  
The bosonic neutral field $\theta_{s}\para{x}$ commutes with every term in the Hamiltonian and therefore can be condensed. After the condensation of $\theta_s$ the $\cos\para{\sqrt{3}\theta_{s}}$ can be replaced with its expectation value. Doing so, we achieve the following effective sine-Gorodn Hamiltonian:
\begin{eqnarray}
\mathcal{H}_{\rm eff} =\frac{1}{4\pi} && \int dx \left[\para{\partial_x \varphi_{c}}^2+\para{\partial_x \theta_{c}}^2\right]\cr
-u&& \int dx ~\left[\Delta_{\perp} \cos\para{\sqrt{3} \varphi_{c}}+t_{\parallel} \cos\para{\sqrt{3}\theta_{c}} \right].
\end{eqnarray}
where $u=4\braket{\cos\para{\theta_s}}$. Since the conformal dimension of the cosine perturbations is $(3/2)$ the above effective Hamiltonian can be identified with the
well studied $\beta^2=6\pi$ self-dual sine-Gordon model according to the notation of Ref. \cite{lecheminant2002}. In Ref. \cite{vaezi2013c} we have shown that the $\beta^2=6\pi$ sine-Gordon model describes the 
low energy physic of a $Z_3$ parafermion chain. This immediately suggests that the self-dual point i.e. $t_{\parallel}=\Delta_{\perp}$ is a critical point with $Z_3$ parafermion CFT description. Ref. \cite{lecheminant2002} presents two two other proofs for this result. The resulting $Z_3$ parafermion CFT has a chiral (anti-chiral) sector with $c=4/5$ ($c=-4/5$) central charge. It has six different quasiparticles, three of which are Abelian anyons: $\mathbb{I}, \psi, \psi^\dag$, where $\psi$ is the parafermion primary field. The remaining three primary field are non-Abelian excitations: $\sigma, \sigma^\dag, \epsilon$, where $\sigma$ is the spin field and $\epsilon=\sigma \psi$ is the Fibonacci anyon. The quantum dimension of these non-Abelian excitations is $\frac{1+\sqrt{5}}{2}$. 

Next, we follow the method developed in \cite{mong2013,teo2011,vaezi2013} and consider a 2D array of the 1D chains with proper inter-chain couplings. Each of the 1D chains consists of a pair of counter-propagating $Z_3$ parafermion
modes. Now we can consider two different possibilities. In the first scenario we can couple the parafermion modes such that every parafermion mode is gapped except for the right-moving parafermion on the topmost chain and the left-moving parafermion mode on the bottom-most chain (see Fig. \ref{fig:Fig5}). In the second scenario the topmost left-moving and bottom-most right-moving parafermion modes remain gapless and all other parafermion modes gap out.

In addition to the remaining chiral parafermion modes,  the two outer-most edge modes of the sample are still chiral with $c=2$ central charge. These two bosonic modes correspond to the edge between the parent $(221)$ state and vacuum. The two scenarios sketched above result in an additional $Z_3$ parafermion mode whose central charge
is $c = 4/5$ in the first scenario and $c=-4/5$ in the second one. Hence, the total chiral central charge of the state is $c = 2\pm 4/5$. The first scenario gives rise to the Fibonacci phase with $c=14/5$ central charge with only two primary fields:  identity operator and Fibonacci anyon. In Ref. \cite{vaezi2014} we have studied a related problem and we obtain $c=14/5$ from two other methods. So we believe that the first scenario that yields the Fibonacci phase is more likely. 

%
\end{document}